\documentclass[usenatbib]{mn2e}
\usepackage{times}
\usepackage{epsfig}
\usepackage{threeparttable}
\newcommand{\kms}{~km~s$^{-1}$}

\def \etal      {{\it et al.\ }}

\def \eg        {{\it e.g.,\ }}
\def \Mr        {{\rm\ M_{R}}}
\def \Mb        {{\rm\ M_{B}}}

\begin{document}
\title[Bimodal luminosity functions in galaxy groups]
{The Group Evolution Multiwavelength Study (GEMS): bimodal luminosity functions
in galaxy groups}
\author[Trevor A. Miles et al.]{%
       \parbox[t]{\textwidth}{Trevor A. Miles$^{1}$\thanks{E-mail:
tm@star.sr.bham.ac.uk},
Somak Raychaudhury$^{1}$,
 Duncan A. Forbes$^{2}$, Paul Goudfrooij$^{3}$, Trevor~J.~Ponman$^{1}$ \& 
Vera Kozhurina-Platais$^{3}$}
\vspace*{6pt} \\
$^{1}$School of Physics and Astronomy, The University of Birmingham,
Birmingham B15 2TT, UK\\
$^{2}$Centre for Astrophysics and Supercomputing, Swinburne University,
Hawthorn, VIC 3122, Australia\\
$^{3}$Space Telescope Science Institute,       
3700 San Martin Drive, Baltimore, MD 21218}
\date{
MNRAS accepted - 2004 August
}
\pagerange{\pageref{firstpage}--\pageref{lastpage}}
\pubyear{2004}
\label{firstpage}
\maketitle
\begin{abstract}
We present $B$ and $R$-band luminosity functions (LF) for a sample of
25 nearby groups of galaxies.
We find that the LFs of the groups with low X-ray
luminosity ($L_X\! <\! 10^{41.7}$ erg~s$^{-1}$) 
are significantly different from
those of the X-ray brighter groups, showing a prominent dip around $\Mb
\!=\!-18$. While both categories show lack of late-type galaxies in
their central regions, X-ray dim groups also show a more marked
concentration of optical luminosity towards the centre. A toy
simulation shows that in the low velocity dispersion environment, as
in the X-ray dim group, dynamical friction would facilitate more rapid
merging, thus depleting intermediate-luminosity galaxies to form a few
giant central galaxies, resulting in the prominent dip seen in our
LFs.  We suggest that X-ray dim (or low velocity dispersion) groups
are the present sites of rapid dynamical evolution rather than their
X-ray bright counterparts, and may be the modern precursors of fossil
groups. We predict that these groups of low velocity dispersion would
harbour younger stellar populations than groups or clusters with
higher velocity dispersion.

\end{abstract}


\section{Introduction}
Galaxy luminosity functions (LF) provide a means of comparison between the
populations of galaxies of various luminosities in
different environments, and contain valuable information about the
physical processes that feature prominently in galaxy formation.
Models of galaxy formation have been related to observable luminosity
functions of galaxy populations from the basic 
principles of \cite{whiterees78} to more recent detailed models
including stellar evolution, merging, cooling and feedback
\citep{wf92, cole94, somerville99, kauff99, benson03}. These models
are now being put to the test by high-quality photometric observations
of wide-field and deep samples of the field and in highly clustered
regions.

Early researchers, such as \citet{tg76}, motivated by the potential
usefulness of the LF as a distance indicator, accepted the paradigm
that all ensembles of galaxies, from groups to rich clusters, follow a
Universal luminosity function.  The shape of this luminosity function
is usually modelled as \citep{schechter76}
\begin{equation}
\phi(L)=(\phi^{\ast}/L^{\ast})(L/L^{\ast})^{\alpha}e^{-L/L^{\ast}}.
\label{eq:schechter}
\end{equation}
The Schechter function drops sharply at bright magnitudes
and rises at the faint end following a power law of slope
$\alpha$, the transition occurring around the luminosity
$L^{\ast}$, ${\phi^{\ast}}$ being the normalisation parameter. 
Field samples usually yield $\alpha \!=\! -1.0$ (\eg APM-Stromlo,
\citealt{apmstromlo95}, 
$b_J$-band), but shallower or steeper slopes have been found (\eg\
LCRS, \citealt{lcrs96}, $R$-band, $\alpha \!=\! -0.7$; 2dF,
\citealt{2df-lf}, $b_J$-band, $\alpha \!=\! -1.19$).

However, accurate photometry of the faint end of the LF reveals that
the LF may be more complex, and features such as dips at intermediate
luminosities or an excess of faint galaxies are sometimes found,
making a single Schechter function a poor fit to the data.
\cite{huns98} find a dip at $\Mr = -18$ in the composite LF of 37
Hickson compact groups (HCG), an effect which has been found at
brighter magnitudes in a few rich clusters of galaxies (e.g. Abell
2554, Coma) by various other studies \citep[e.g.][]{tt02,smith97}.
Indeed, there has been a recent move towards modelling the LF of
brighter galaxies ($M_B<-18.5$) as Gaussian, and that of the fainter
ones with a Schechter function \citep[e.g.][]{lobo97, ap00}.  This
approach seeks to explain the peaks and dips in the LF as due to a
varying mix of galaxies of different morphological types in different
environments, and highlights the connection between the evolution of
galaxies and their local environment
\citep[e.g.][]{fergsan91,jerjen-lf}.

Furthermore, there has been a lack of consensus about the overall
shape of the LF in groups of galaxies. \cite{zabmul00} found that the
LF of X-ray bright groups shows an excess of faint galaxies, with a
Schechter index of $\alpha=-1.3\pm 0.1$, while \cite{doh91} find
evidence of depletion of faint galaxies $\alpha \!=\!  -0.2 \pm 0.1$
in HCGs.  \cite{zepf91} find an intermediate value of $\alpha \!=\!
-1.0 \pm 0.1$, similar to that in the field, in 17 HCGs, while
\cite{huns98} find a slight excess of dwarf galaxies when compared to
the field ($\alpha \!=\! -1.17\pm 0.1$) in 37 HCGs.  A possible source
of the anomalies between these studies is the subtraction of the
extragalactic background, which is sensitive to inhomogeneities in the
large-scale structure, and lack of information of group membership,
which is in principle redeemable with redshift measurements.

Here we suggest that where sufficient redshift and colour information
is available, categorising the LF of groups of galaxies in terms of
X-ray luminosity (or equivalently, group velocity dispersion, \eg\ 
\citet{hp2000}) is an effective method for investigating the
connection between differences in the shapes of group luminosity
functions and the local environment, and probing the underlying galaxy
populations and history of evolution.

In the following section, we present observations of 25 groups of
galaxies, and in \S3, compute their luminosity function in the $B$
and $R$-band,
splitting the sample in categories based on their X-ray luminosity.
In \S4, we use toy-model simulations to qualitatively explain the
differences in the LFs of X-ray bright and dim groups, and in \S5,
we discuss the implications in the context of the evolution of galaxies
in groups. We have used  
$H_0=70$ \kms Mpc$^{-1}$; q$_0=0.5$ throughout.

\section{Observations and Analysis}

Our sample of 25 groups is drawn from the Group Evolution
Multi-wavelength Study \citep[GEMS]{osmond04} of sixty groups of
galaxies, compiled to incorporate a wide variety of groups
representing a wide range of evolutionary stage and local environment.
The master sample was compiled by cross-correlating a list of over
4000 catalogued groups with archival ROSAT PSPC X-ray observations
with integration $>$10~ks. A large fraction of these were detected in
the X-ray, and for the others we have upper limits for their X-ray
flux.

\begin{table*}
\begin{threeparttable}
\caption{The sample of groups of galaxies used in this study\label{table1}}
\begin{tabular}{lccccrrcr}
\hline
Group & R.A.  & Dec & Obs.\tnote{a}
&Distance\tnote{b}
&$\sigma$ 
& $\log L_X$
& Emission\tnote{d}\\
& J2000  & J2000 & &Mpc & km/s & ergs/s  \tnote{c} & Type \\ \hline
NGC  524 & 01:24:47.8 & +09:32:19 &I  & 35.4 & 175  & 41.05  & Galaxy  \\
HCG   10 & 01:26:07.4 & +34:41:27 &I  & 68.2 & 231  & 41.70  & Group  \\
NGC  720 & 01:53:00.4 &--13:44:18 &E  & 23.2 & 273  & 41.20  & Group   \\
NGC 1052 & 02:41:04.8 &--08:15:21 &I  & 20.3 & 91   & 40.08  & Galaxy   \\
HCG  22  & 03:03:31.0 &--15:41:10 &E  & 38.7 & 25   & 40.68  & Group    \\
NGC 1332 & 03:26:17.1 &--21:20:05 &E & 22.9 & 186  & 40.81  & Galaxy    \\
NGC 1566 & 04:20:00.6 &--54:56:17 &E & 20.8 & 184  & 40.41  & Galaxy   \\ 
NGC 1587 & 04:30:39.9 & +00:39:43 &I  & 55.2 & 115  & 41.18  & Group     \\
NGC 2563 & 08:20:35.7 & +21:04:04 &I  & 73.5 & 384  & 42.50  & Group     \\
NGC 3227 & 10:23:30.6 & +19:51:54 &I  & 26.5 & 169  & 41.23  & Galaxy     \\
NGC 3396 & 10:49:55.2 & +32:59:27 &I  & 31.2 & 106  & 40.53  & Galaxy     \\
NGC 3607 & 11:16:54.7 & +18:03:06 &I  & 23.5 & 280  & 41.05  & Group     \\
NGC 3640 & 11:21:06.9 & +03:14:06 &I  & 28.5 & 211  &\llap{$<$}40.37&Undetected\\
NGC 3665 & 11:24:43.4 & +38:45:44 &I  & 37.2 & 87   & 41.11   & Group    \\
NGC 4151 & 12:10:32.6 & +39:24:21 &I  & 23.0 & 102  &\llap{$<$}40.20 & Undetected\\
NGC 4261 & 12:19:23.2 & +05:49:31 &I  & 41.2 & 197  & 41.92   & Group   \\
NGC 4636 & 12:42:50.4 & +02:41:24 &I  & 10.3 & 284  & 41.49   & Group   \\
NGC 4725 & 12:50:26.6 & +25:30:06 &I  & 25.1 & 49   & 40.63   & Galaxy   \\
NGC 5044 & 13:15:24.0 &--16:23:06 &I  & 33.2 & 426  & 43.01   & Group    \\
NGC 5322 & 13:49:15.5 & +60:11:28 &I  & 34.9 & 166  & 40.71   & Galaxy  \\
HCG   68 & 13:53:26.7 & +40:16:59 &I  & 41.1 & 191  & 41.52   & Group     \\
NGC 5846 & 15:06:29.2 & +01:36:21 &E  & 29.9 & 346  & 41.90   & Group     \\
NGC 7144 & 21:52:42.9 &--48:15:16 &E & 26.6 & 41   & 40.33  & Galaxy   \\
HCG   90 & 22:02:08.4 &--31:59:30 &E & 36.3 & 131  & 41.49   & Group     \\
IC  1459 & 22:57:10.6 &--36:27:44 &E & 25.6 & 223  & 41.28   & Group    \\
\hline
\end{tabular}
\begin{tablenotes}
\item[a] Observations obtained at (I) the INT, La Palma and (E) the 2.2m ESO/MPI telescope  
\item[b] Distance measured from the redshift
of central galaxy (assuming $H_0=70$ \kms Mpc$^{-1}$; q$_0=0.5$),
corrected for local bulk flows.
\item[c] Bolometric X-ray luminosity from \citet{osmond04}
\item[d] ``Group'' emission refers to unambiguous 
detection of diffuse
X-ray emission from hot gas in the group potential, not belonging
to any galaxy. Those marked ``Galaxy''
have no discernible diffuse emission that doesn't belong to
individual galaxies, and those ``undetected'' fall below
the detection threshold of the ROSAT PSPC observation used.
\end{tablenotes}
\end{threeparttable}
\end{table*}

\subsection{Photometry and calibration} 

Of this sample, 17 groups were observed at the 2.5m Isaac Newton
telescope at the Roque de Los Muchachos Observatory, La Palma, between
2000 February 4--10.  The detector used was the prime focus Wide Field
Camera (WFC), which is an array of four thinned EEV CCDs, each with an
area of $2048\times 4096$ pixels, each 0.33~arcsec across. Each CCD
thus can image an area of $22.5\times 11.3$ arcmin of sky, together
covering 1017 sq.arcmin. Images obtained with broadband BVI filters
were processed using standard NOAO IRAF packages.

A further 8 groups were observed with the 2.2m ESO/MPI telescope at
La~Silla observatory using the Wide Field Imager (WFI), between 2001
August 7--10. The WFI is a focal reducer-type mosaic camera with 8 CCD
chips each with $2048\times 4098$ pixels, the pixel scale is
0.238~arcsec providing a field of view for the whole camera of
$34\times 33$ arcmin. Photometry was performed with broadband BRI
filters.  Exposures of standard star fields \citep {landolt92} were
taken during both observing sessions. The ESO images were reduced
using the ESOWFI package implemented within IRAF \citep{jv2000}.

The average integration times used per image were 780s in the B-band,
390s in the R-band and 290s in the I-band for the INT run, and 600s in
the B-band, 300s in the R-band and 180s in the I-band for the ESO run.
The median seeing achieved throughout the run was between 1.2 arcsec
FWHM in the $B$ filter, 1.1 arcsec in $R$ and 1.0 arcsec in $I$
measured from the brightest unsaturated stars in the field for INT
observations and 1.3 arcsec in the $B$ filter, 1.1 arcsec in $R$ and
0.9 arcsec in $I$ for ESO observations.

\begin{figure}
\epsfig{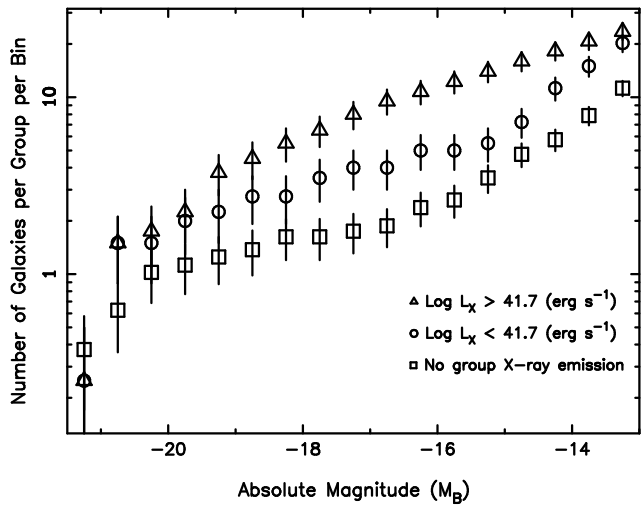}
\caption{The Cumulative $B$-band 
Luminosity Function for 25 groups of galaxies
separated into three categories: X-ray bright groups
(triangles), X-ray dim (circles) and X-ray undetected groups (those
with no discernible group emission, squares). It is clear that
the LF of the X-ray bright groups is different from that of the 
other two classes, the latter showing a ``dip''
between $-19\!<\!M_B\!<\!-17$. We combine the X-ray dim and undetected
groups into a single category in the rest of the paper.}
\label{fig:cumul3}
\end{figure}

\subsection{Galaxy Selection}
Images were identified and fluxes extracted using the SExtractor
package \citep{sext96}. For all identified objects, positions,
magnitudes, fluxes, star/galaxy classifiers and flags were
written to a catalogue. The ``stellaricity'' parameter uses a neural
network-based algorithm to classify images.  Detections were checked
visually and objects with stellaricity $>$ 0.9 were deemed to be
definitely stellar and therefore not subject to further analysis. All
objects with FWHM less than the PSF were discarded as noise. All 
galaxies identified by this procedure were visually inspected, and
classified as early (E/S0) or late (S/Irr) types.

A fixed aperture, set to be slightly greater than the seeing, was used
to obtain magnitudes in all filters for colour selection. Objects in
different filters were matched and aperture magnitudes were subtracted
to find colours.

Galaxies were selected as being likely group members on the basis of
their $(B-R)$ colour.  A cut-off was selected at $(B-R) \!=\! 1.7$ for
all groups, which represents a conservative selection criterion and
removes the majority of background galaxies.  This selection criterion
was chosen on the basis of the work of \cite{fuku95}, where, using
colours of galaxies at various redshifts, obtained in 48 photometric
bands, it was shown that all elliptical galaxies with $(B-R) \!>\!
1.7$ have a redshift greater than 0.2. The most distant
group in our present sample has a redshift of $z \!=\! 0.016$ (NGC
2563 group).

Distances to the groups were calculated from redshifts, by estimating
peculiar velocities according to a model of the local velocity
field, including the infall into Virgo and the Great Attractor
(for further details, see
\citet{osmond04}).

\subsection{Background Subtraction}

Given that the number of galaxies belonging to each group is small, a
major reason for disagreement between various authors as to the nature
of the galaxian LF in groups is that the number of background and
foreground galaxies seen in each group in projection can be rather
significant.

\cite{flint01} point out that while a system of classification 
by colour, like ours,
is appropriate for the majority of galaxy types, there
exists a population of very red ($B\!-\!R>1.7$), low surface
brightness galaxies which may be misclassified as background objects.
On the other hand, high-resolution HST photometry of the elliptical
galaxies IC4051 and NGC4481 in Coma \citep{andreon02} has revealed
blends of globular clusters that were classified as single extended
sources from the ground, which caused significant contamination of the
faint end of the luminosity function in the Coma cluster at $R \ga
21$.  However, since the average distance modulus of our groups is
$\sim$33 (as opposed to 35 for Coma), and we compute LFs down to
$M_B\!=\! -13$, we expect contamination from blends of globular
clusters to be minimal.
\begin{figure*}
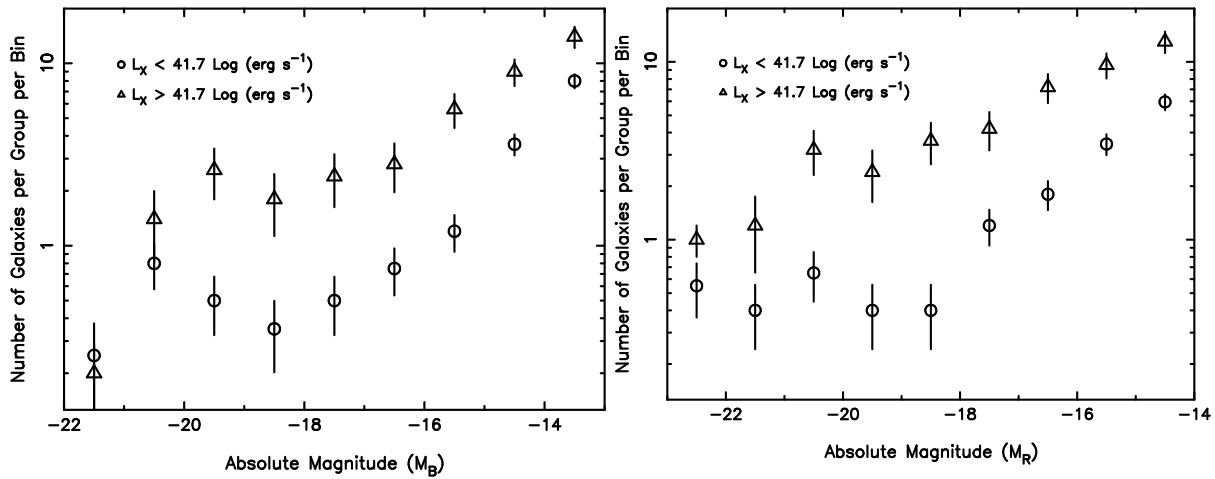

\centerline{
\epsfig{file=tm2.ps,angle=-90,width=0.45\hsize}
\epsfig{file=tm3.ps,angle=-90,width=0.45\hsize}
}
\caption{Differential $B$-band (left) and 
$R$-band (right) Luminosity Function of 
25 GEMS Groups of galaxies: X-ray bright groups ($L_X>10^{41.7}$ erg~s$^{-1}$,
triangles) and  X-ray dim groups ($L_X<10^{41.7}$ erg~s$^{-1}$, circles).
It is clear that
the LF of the dim groups show a ``dip'' in
the LF between $-19<M_B<-17$ and $-20<M_R<-18$. The bright groups
too show dips, albeit not so pronounced, in the same intervals.
\label{fig:difflf2}}
\end{figure*}

Having tried various processes of background subtraction, we chose to
evaluate the luminosity function of all galaxies with $B\!-\!R<1.7$
from the regions outside a radius of $R_{500}$ from the centre of the
group (values from \citealt{osmond04})
and use it as the background for subtraction.

\section{The Luminosity function of galaxies in groups}

The galaxian luminosity function (LF) of the groups is evaluated here
by co-adding galaxies of the same absolute luminosity of several
groups, since the number of member galaxies in each individual group is small.
The groups in our sample represent very diverse systems in terms of
their content and physical properties. We therefore chose to assemble
the galaxies according to the X-ray luminosity of their parent groups,
which provides a measure of the mass and velocity dispersion of
the group, and helps to
distinguish between virialised systems and dynamically young and
forming systems. We use the X-ray luminosities measured by
\citet{osmond04}, who used ROSAT PSPC observations in the 0.5--2 keV
range, fitting $\beta$-profiles after point source removal,
extrapolated to estimate the bolometric X-ray luminosity.

We characterised the parent groups as 
X-ray bright if their bolometric
X-ray luminosity is more than the median of the sample, $L_X=10^{41.7}$
erg~s$^{-1}$, and X-ray dim if less. This X-ray luminosity refers to
that of the group plus any central galaxy that might
exist. In addition, we add a third category (``X-ray undetected''
groups) where there is no discernible group emission, and all the
diffuse emission, if any, can be accounted for by emission from
individual non-central galaxies in the group.

The cumulative galaxian Luminosity Function ($B$-band) of the 25 GEMS
Groups in our sample, separated into the three categories described
above (X-ray bright, dim and undetected), and plotted for all galaxies
together in each category, is shown in Fig.~\ref{fig:cumul3}.  A clear
difference between the X-ray bright groups and those in the other two
categories is immediately evident, with a depletion or `dip' in the
number of galaxies with magnitudes between $-19\!<\!{\Mb}\!<\!-17$ for
the latter.  In the cumulative LF, as is plotted in
Fig.~\ref{fig:cumul3}, this shows up as a flat region in the above
magnitude range.

The possibility that this dip is due to some of the galaxies being
systematically missed can be ruled out, as we are complete to over
five magnitudes dimmer than the position of the dip.  According to the
Poisson statistics used to calculate the error bars, we can also
reject the hypothesis that a statistical fluctuation accounts for the
differences between the groups.

\subsection{X-ray bright and dim groups}

There is no significant difference, as revealed by the
Kolmogorov-Smirnov (KS) test, between the shape of the luminosity
functions of the X-ray dim and the X-ray undetected groups
(circles and squares
in Fig.~\ref{fig:cumul3}).  Therefore, for the rest of the paper we will
combine them into a single category of X-ray dim groups.

\begin{figure}
\epsfig{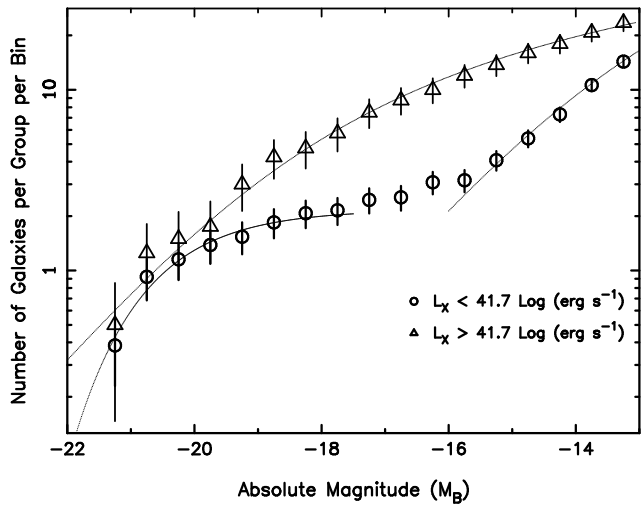}
\caption{
Cumulative $B$-band Luminosity Function of 25 GEMS Groups of galaxies
grouped into the same categories as \ref{fig:difflf2} The former is
fit with a single Schechter function, whereas the superposed curve on
the latter represent two Schechter functions.
\label{fig:complf2}}
\end{figure}

The LFs of the groups of Fig.~\ref{fig:cumul3} are now re-plotted into
two categories of X-ray bright ($L_X\!>\! 10^{41.7}$ erg~s$^{-1}$) and
dim (those that have lower X-ray luminosity).  The differential
$B$-band and $R$-band luminosity functions are plotted in
Fig.~\ref{fig:difflf2}.  The null hypothesis that the two
distributions were drawn from the same population was rejected at the
99.999$\%$ level of confidence for both LFs shown in
Fig.\ref{fig:difflf2}.  These LFs clearly show the position of the dip
for the X-ray dim groups, between $-19\!<\!M_B\!<\!-17$ and $-20\!<\!
M_R \!<\! -18$ respectively.  Interestingly, the X-ray bright group
LFs also show a slight dip in the same interval, though by far not as
prominently, such that it is not easily apparent from the
corresponding cumulative LF ($B$-band only, Fig.~\ref{fig:complf2}).

\begin{figure*}
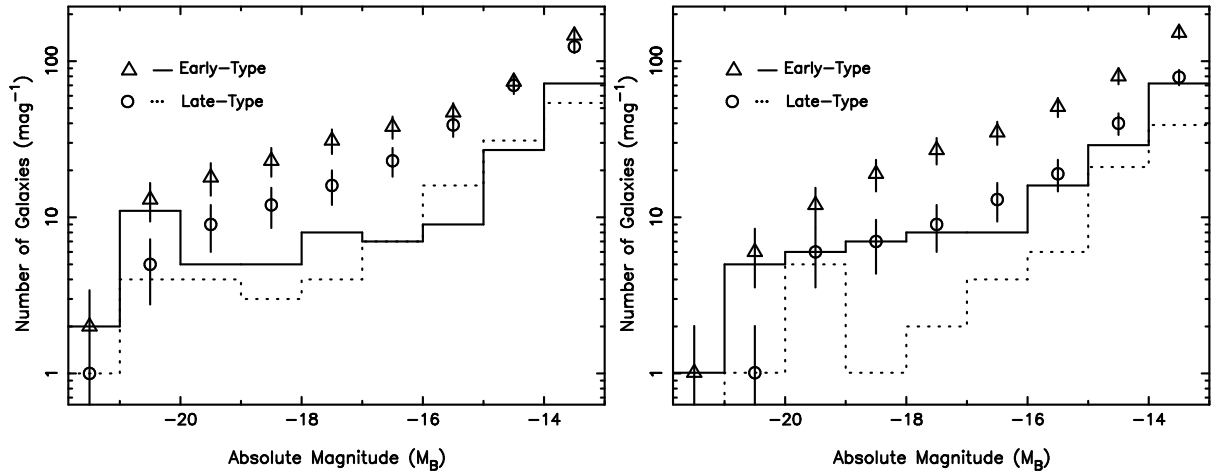

\centerline{\epsfig{file=tm5.ps,angle=-90,width=0.45\hsize}
\epsfig{file=tm6.ps,angle=-90,width=0.45\hsize}}
\caption{$B$-band
luminosity functions 
for early and late-type galaxies for the X-ray dim (left) and bright
(right) groups. The histograms show the differential LFs (solid: early,
dotted: late), whereas the points represent the cumulative
LFs (triangles: early, circles: late).
\label{fig:morph-hist}}
\end{figure*}

We fit the luminosity function of X-ray bright groups with a single
Schechter function of the form (\ref{eq:schechter}), yielding best fit
values in the B-band of $M^\ast = -20.1 \pm 0.1$ and $\alpha = -1.04
\pm 0.01$, with the reduced $\chi^2$ for the fit being 0.9.  It is
clear that the LF of the X-ray dim groups in Fig.~\ref{fig:difflf2} is
not well-fit by a single smooth function: an attempt to fit it by a
single Schechter function yielded a reduced $\chi^2$ of 54. If
we fit the LF in this category with two Schechter functions, we
obtain, for the bright end, $M^\ast = -20.5\pm 0.1$ and $\alpha =
-0.04 \pm 0.01$, whereas for the faint end, the best fit yields
$M^\ast = -14.0 \pm 0.1$ and $\alpha = -1.01 \pm 0.01$.

\begin{figure}
\begin{center}
\epsfig{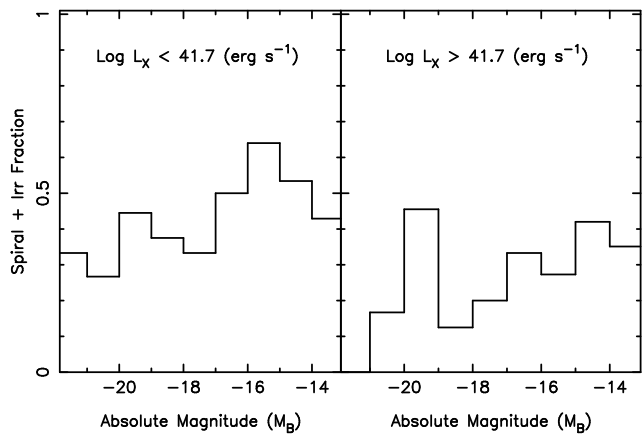}
\caption{The fraction of late-type galaxies  as a function of absolute
magnitude, for the X-ray dim (left) and bright (right) groups.
The X-ray dim groups have a lower fraction of late-type galaxies
at the bright end. 
\label{fig:spfrac}}
\end{center}
\end{figure}

Fig.~\ref{fig:morph-hist} shows the differential and cumulative
$B$-band LFs separately for early-type and late-type galaxies, for
both X-ray bright and faint parent groups. The LF of early-type
galaxies  shows a significant dip in the X-ray dim systems, where
there is none for the brighter systems (right panel). Even
though the LF of
late-type galaxies
 seems to have a peak at the bright end for
the X-ray bright groups, the small numbers involved 
(Fig.~\ref{fig:spfrac}) make
the errors on each point considerably large and the brighter
end of the late-type LF rather unreliable for these systems.

\subsection{The radial distribution of galaxy light}

\begin{figure}
\begin{center}
\epsfig{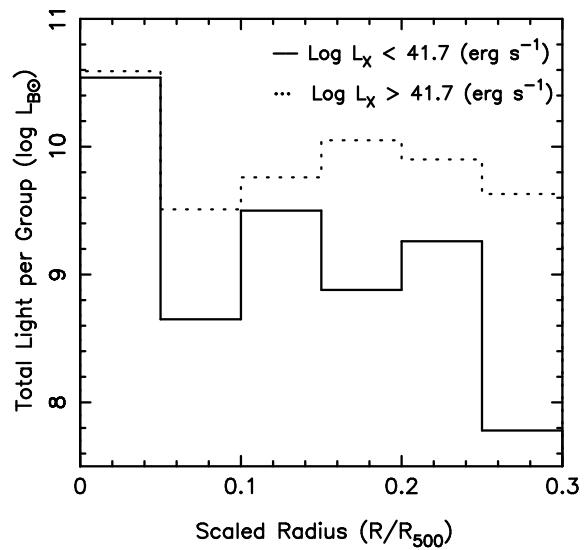}
\caption{
The mean $B$ luminosity of all galaxies in high $L_X$ and low $L_X$ groups
as a function of radius (expressed in terms of the standard radius
$R_{\rm 500}$, determined from the group temperature).  The luminosity
within the same scaled projected radius $R/R_{500}=0.3$ is lower in X-ray dim
groups (solid histogram) than in X-ray bright groups (dotted
histogram), suggesting that the latter are closer to a stage of
virialisation.
\label{fig:meanlight}}
\end{center}
\end{figure}

Having looked at the relative distribution of galaxies of different
luminosities in the group as a whole, we now turn to the distribution
of light as a function of distance from the centre of each group.  The
centre and overdensity radius $R_{500}$ for each group are taken from
\citet{osmond04}. The total blue light within the same scaled 
(projected) radius
$R/R_{500}=0.3$ is lower, and more centrally peaked,
in X-ray dim groups than in X-ray bright
groups (Fig.~\ref{fig:meanlight}), suggesting that 
the groups in the former category are dynamically 
evolving.

\begin{figure}
\begin{center}
\epsfig{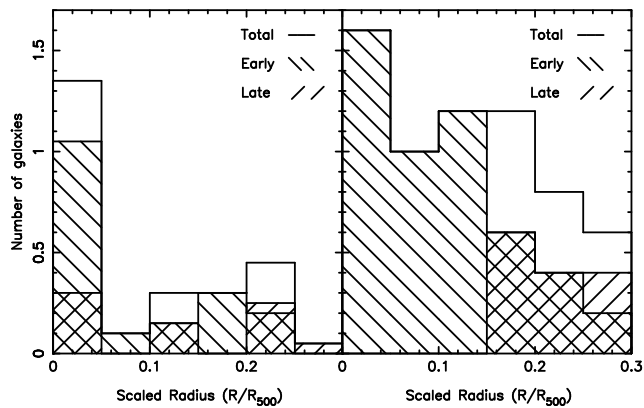}
\caption{Average number of bright ($\Mb\!<\!-17$) early and late-type
galaxies,
plotted as function of scaled projected 
radius, in (left panel) X-ray dim groups
and (right panel) X-ray bright groups respectively.  Both categories
of group show a distribution that is centrally concentrated, and a
lack of late-type galaxies
in the central regions. However, the X-ray dim groups
display a more centrally peaked distribution of early-type
galaxies than their X-ray bright counterparts.
\label{fig:low-highell}}
\end{center}
\end{figure}

The average number of bright ($\Mb\!<\! -17$) early and late-type
galaxies is plotted as a function of radial distance in
Fig.~\ref{fig:low-highell}.  Both the X-ray bright and dim groups show
a distribution of early-type galaxies that is centrally concentrated,
though the X-ray dim groups display a far more centrally peaked
distribution of galaxies than their X-ray bright counterparts.  A KS
test shows that the null hypothesis of the two distributions being
drawn from the same sample is rejected at the 99.98$\%$ level of
confidence.  This is consistent with the observation that late-type
galaxies tend to avoid the central regions of both kinds of groups, in
addition to being rare in X-ray bright ones (as seen in
Fig.~\ref{fig:spfrac}).

\subsection{Comparison with the Coma Cluster and the Leo I group}

 To compare our LFs with
those of other studies, photometric images were obtained of
the Coma Cluster and the Leo~I group, during the same INT observing run
as the Northern GEMS groups and with the same instrumental setup, 
in 2000 February.  

\begin{figure}
\epsfig{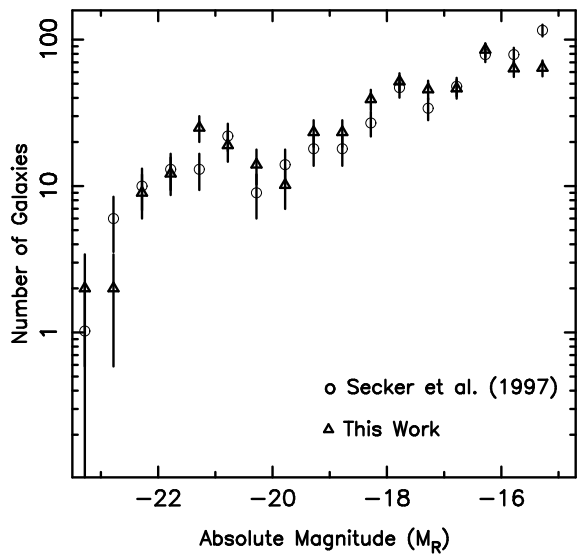}
\caption{Differential
$R$-band Luminosity Function for the Coma cluster. 
Our results are plotted alongside Secker et al. (1997)
to allow direct comparison. The LF of Coma is similar to that
of our X-ray bright groups, though the dip occurs in
a brighter magnitude range.
\label{fig:comalf}
}
\end{figure}

We plot the differential $R$-band luminosity function for the Coma
cluster, following colour selection of member galaxies as above, in
Fig.~\ref{fig:comalf}.  The background subtraction employed here is a
scaled (by relative area) version of that used by \cite{secker97}, to
facilitate direct comparison. The result is consistent with the
\cite{secker97} LF within the errors. Small differences can be
explained by the fact that we use different regions of the Coma
cluster (which could be significant, as found by
\cite{beij02}), 
and the membership criteria applied are slightly different.

Our Coma LF looks reassuringly similar to that of \cite{secker97}, and
reveals a shallow dip reminiscent of the X-ray bright GEMS groups,
albeit in the slightly brighter absolute magnitude range
$-20\!<\!M_B\!<\!-18$. Similar dips have been observed in other
individual rich clusters (e.g. Abell 2554, Smith et al. 1997), though
it does not seem to be a common property of the luminosity functions
of rich clusters, as is apparent from the composite LF of clusters in
the 2dFGRS \citep{2df-clust-lf}.  With an X-ray luminosity of $L_X =
7.3\times10^{44}$ erg/s \citep{Ebeling98}, the Coma cluster is a
couple of orders of magnitude more luminous than our brightest GEMS
group (see Table~1).

\begin{figure}
\begin{center}
\epsfig{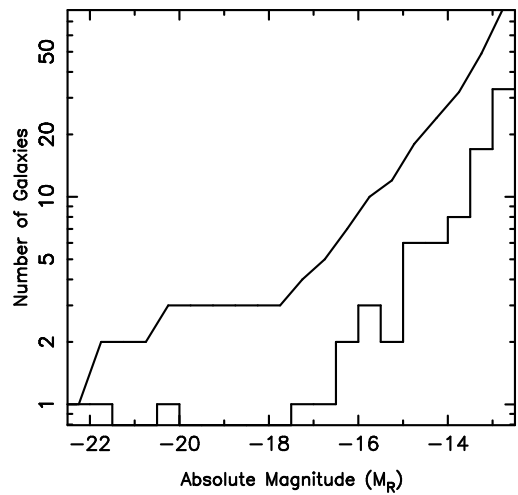}
\caption{$R$-band Luminosity Function of the Leo I group. 
Both differential and cumulative LFs are plotted.  The LF of Leo~I,
which hasn't been detected in the X-ray, is similar to that of our
X-ray dim groups.
\label{leolf}}
\end{center}
\end{figure}

The Leo I group is a well-known nearby ($v\!\sim\! 950$\kms) group of
galaxies, its proximity means that it has received attention from many
researchers, notably \cite{fergsan91}. More recently, \cite{flint03}
have probed the faint end of the $R$-band LF of this group down to $M_R
\sim -10$. Fig.~\ref{leolf} shows our determination of the LF of
Leo~I, which agrees well with both of the above studies.

Perhaps the most striking feature of the Leo I group LF is the lack of
intermediate magnitude galaxies. We fit the cumulative LF with
two Schechter functions, as above for X-ray dim groups, 
obtaining
   $M^\ast = -14.0 \pm 0.1$ and
   $\alpha = -1.20 \pm 0.01$ for the faint end ($M_R > -18$).
This seems to be a rather extreme analogue of the LFs 
of the GEMS X-ray dim groups and indeed no group X-ray emission has been 
detected in the Leo I group.

\section{Galaxy mergers and group evolution} 
\label{sec:toy}

Mergers play an important role in the evolution of galaxies in groups.
Once a group has formed from gravitational collapse, violent
relaxation no longer plays an important role, and the system relaxes
thereafter predominantly through two-body interactions.  Dynamical
friction causes galaxies to fall towards the centre of the group, and
together with time-dependent tidal forces, help to redistribute the
ordered orbital kinetic energy, allowing the galaxies to merge.  Since
the deceleration due to dynamical friction is inversely proportional
to the square of the relative velocity of the interacting galaxies, it
is more likely to be effective in those groups with low velocity
dispersion.


Here we have categorised groups according to their X-ray luminosity,
since there are too few known galaxy redshifts for most of the sample
groups to allow reliably measuring 
their velocity dispersion. Since there
is a power-law relation between $L_X$ and $\sigma$ \citep{hp2000}, we
have assumed our X-ray dim groups to be of lower velocity dispersion
than the X-ray bright groups. Qualitatively, therefore, we can suggest
a link between merger-driven evolution and the velocity dispersion of
the group, to explain the development of the bimodal shape of the LF
in X-ray dim groups, where mergers are more likely.

\begin{figure}
\epsfig{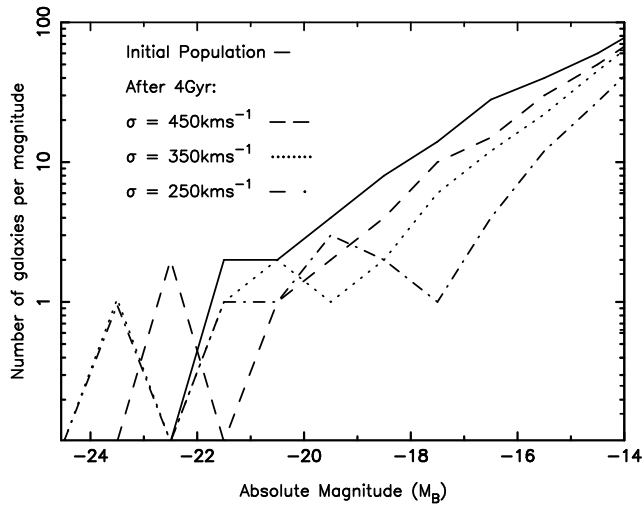}
\caption{The shape of the differential luminosity function of the group,
according to our toy model described in Section~\ref{sec:toy}, after a
period of 4~Gyr, starting from 50 galaxies distributed according to
the luminosity function shown (solid line), for groups of
one-dimensional velocity dispersion $\sigma_e=$250, 350 and 450 km/s.
Note that a dip appears at intermediate luminosities
($-18\!<\!M_B\!<\!-16$) for the lowest $\sigma_e$ group.
\label{fig:4gyrs}}
\end{figure}

To examine the role of mergers in the long-term evolution of the group
luminosity function more quantitatively, we construct a toy model of a
group, starting with 50 galaxies with spherical haloes, drawn from the
initial luminosity function shown in Fig~\ref{fig:4gyrs}, assumed
evenly distributed within a radius of $R\!=\!1$~Mpc. A fixed
mass-to-light ratio is assumed, so that mass scales simply with
luminosity. The distribution of mass within each galaxy is assumed to
follow a Plummer model, and the internal velocity dispersion of each
galaxy is calculated from the Faber-Jackson relation $\sigma_i=220
(L/L^\ast)^{0.25}$ km/s \citep{bm98}.

We follow \cite{mh97} in calculating the rate of merger
of pairs in this ensemble, which varies as
$R_V^{-3}\,r_h^2\,\sigma_i^4\,\sigma_e^{-3}$, where $r_h$ is the
half-mass radius of each galaxy (taken to be $0.77\,R_V$
for a Plummer model), and $\sigma_i$
and $\sigma_e$ the one-dimensional velocity dispersion of each galaxy
and the group respectively.  This derivation assumes that the group is
in dynamical equilibrium, and that velocities of galaxies within it
follow a Maxwellian distribution. Following the prescription of
\cite{mh97}, we apply a correction factor of 0.25 to 
the merger rate to account for a
finite limit of the binding energy of a galaxy for tidal disruption.

Now here is why we call the model we use a ``toy'' one. Since the
\cite{mh97} merger rate applies to galaxies of identical mass, we
calculate the merger rate separately in each luminosity bin (i.e. we
allow only for mergers of equal mass galaxies). This exercise is
merely meant to examine how fast galaxies of various luminosities
merge, and whether the bimodal nature can develop preferentially in
low-dispersion systems.

As two galaxies merge, the light is re-distributed according to the
galaxy mass. After a period of evolution $T$, the new galaxy
magnitudes are estimated, the LF plotted and the merger rate of the
new population calculated. Fig.~\ref{fig:4gyrs} shows the result after
$T\!=\!4$~Gyrs for three different groups of one-dimensional velocity
dispersion $\sigma_e\!=$250, 350 and 450 km/s respectively.  The lower
the velocity dispersion of the group, the higher the evidence of
evolution, particularly in the formation of a few galaxies at the
bright end of the LF.

\begin{figure*}
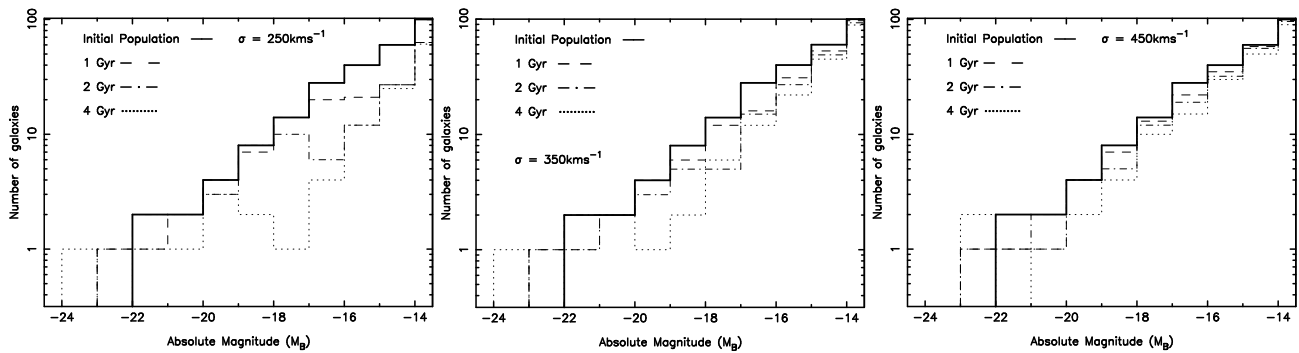

\begin{center}
\epsfig{file=tm13.ps,angle=-90,width=0.32\hsize}
\epsfig{file=tm14.ps,angle=-90,width=0.32\hsize}
\epsfig{file=tm15.ps,angle=-90,width=0.32\hsize}
\end{center}
\caption{The evolution of the luminosity function of a group
of galaxies,
according to the toy model described in Section~\ref{sec:toy},
after 1~Gyr, 2~Gyr and 4~Gyr, for groups with velocity dispersion
$\sigma_e=$250 (Left panel), 350 (Middle) and 450 km/s (Right panel).
The growth of the bright end of the LF is more rapid
in the lower $\sigma_e$ group.
\label{fig:evol-lf}}
\end{figure*}

Fig.~\ref{fig:evol-lf} shows the evolution of the differential LF of
three groups, of one-dimensional velocity dispersion $\sigma_e=$250,
350 and 450 km/s respectively, after time steps $T=$1, 2 and 4~Gyrs,
according to the same model. Even though we do not expect this model
to be a realistic representation of a real group, this simulation does
illustrate an important point. As time goes by, the faint end of the
luminosity function remains featureless, while at the bright end, one
or two very bright galaxies develop in the low-dispersion systems like
the ones illustrated in Fig.~\ref{fig:evol-lf}, at the cost of
intermediate- luminosity galaxies, which leads to the formation of a
dip at these luminosities.  This effect is more prominent for the
lower-$\sigma_e$ systems, where dynamical friction is expected to be
more effective in ensuring that galaxies fall towards the centre of
the group potential, as well as facilitating galaxy interactions and
mergers.

 We also note that in this formalism, only mergers between
equal-mass galaxies are considered.  A more realistic model would tend
to enhance the effect of differentiation between the low-mass and
high-mass galaxies, since the probability of merger between a
high-mass and low-mass galaxy would be higher than that between two
low-mass ones, and the bright end of the LF would be progressively 
enhanced as a result of mergers.

\section{Discussion and Conclusions} 

As the exercise with the toy model in the previous section
illustrates, if one starts with a Schechter luminosity function and
lets a group evolve with time, the LF does not appear to evolve
uniformly as dwarfs merge to form brighter and larger galaxies, but a
number of factors have the potential to alter the total group LF over
time.

Observations of dwarf galaxies in our local neighbourhood,
particularly in the Local Group, shows us that dwarf ellipticals and
spheroidals are preferentially clustered around the brighter galaxies.
Statistical analyses of the distribution of satellites around giant
galaxies \citep{lorrimer94,loveday97} indicate that faint companions
are more strongly clustered about the primary galaxy than their
brighter counterparts. 

The collision time for galaxies in a group is proportional to
$R^{-2}$, where $R$ is the radius of a spherical galaxy.  Galaxies
vary widely in size; for a galaxy with absolute magnitude $M_B=-18$, a
spiral would typically have $R\sim$20 kpc, whereas an elliptical
$R\sim$5 kpc \citep{CO96}.  The faint end of the LF is dominated by
dwarf ellipticals and spheroidals that are less likely to merge with
each other than bright or intermediate luminosity spirals.  Indeed, we
would expect the merger of a dwarf satellite with its parent bright
galaxy to be a more likely event than two dwarfs merging to form an
intermediate luminosity galaxy.
In the Local Group (an X-ray dim group), there is direct evidence of
this happening with satellites of our galaxy. The Sagittarius dwarf
galaxy, for example, is in the process of being accreted by the Milky
Way \citep{ibata94}, and indeed there is evidence of streams across
the sky consisting of remnants of satellites being tidally torn apart
by the Milky Way in the process of eventual mergers
\citep{lblb96}. 

As the low-mass dwarfs merge with the high-mass galaxies, we can
expect the shape of the faint end of the total group luminosity
function to retain its initial shape as the group evolves.  The bright
end of the LF in groups and clusters, on the other hand, can be
expected to be modified as the brightest galaxies grow in luminosity
due to mergers between the brighter galaxies, and the intermediate
luminosity galaxies get depleted. This will happen preferentially in
systems of lower velocity dispersion, where mergers are more frequent,
as evidenced in the observations of X-ray dim groups
(Fig.~\ref{fig:difflf2}). In the field, where mergers are rare, the
intermediate-luminosity dip in the LF is not expected to appear, which
is indeed the case \citep[e.g.][]{2df-lf}.

Further circumstantial support for this hypothesis comes from the
observations of the distribution of light in
groups. Figs.~\ref{fig:meanlight} \& \ref{fig:low-highell} show that
the light of the X-ray dim groups is dominated by one or two central
bright galaxies, and in general the light distribution is more
centrally peaked than in X-ray bright groups. This is what one expects 
from an evolutionary
scenario illustrated in Fig.~\ref{fig:evol-lf}, where 
dynamical friction causes galaxies to fall towards the
the centre of the group, where the rapid merger
of intermediate-luminosity galaxies form the brightest central
galaxies, while the faint end of the LF shows no appreciable change in
slope.

We also find that the X-ray dim groups, where we see the most
prominent dips in the LF, tend to contain a lower fraction of
early-type galaxies, and to have less luminous Brightest Group Galaxies 
\citep[BGG]{khos04}, than the X-ray bright groups.
This appears to imply either (a) many of the early-type galaxies in
the X-ray bright groups did not result from galaxy mergers, or (b) the
merger process which has resulted in the high early-type fraction in
these X-ray bright groups did not produce a LF dip, or (c) the dips in
the LFs of the brighter groups have been filled in in some way. It is
difficult to imagine processes that would lead to (c), and (a) is
rather unlikely given that X-ray bright groups tend to have the most
massive early-type BGGs.

The lack of X-ray emission in some of the X-ray dim systems could well
mean that they are yet to form virialised entities, in which case they
are late-forming systems, which are only now reaching a state of high
density.  The most likely interpretation of the shape of our LFs thus
seems to be that in the X-ray dim systems, most of the galaxy mergers
that are responsible for the dip in the LF have taken place in the
recent past.  In contrast, the X-ray bright, high-$\sigma$ systems are
probably not undergoing much merging at all right now (due to the high
galaxy velocities), so the mergers which have affected their
morphological mix and LF, will have taken place at earlier epochs,
quite probably in a variety of structures which have since merged to
form the group we see now.  The process of merging in a variety of
different precursor systems, over a range of epochs (having a range of
galaxy densities) would lead to a very diverse set of contributions to
the aggregate LF, which would then be unlikely to show a coherent dip.

That rich clusters of galaxies in general do not have bimodal LFs
of this kind is clear from the composite cluster LFs found by
\citet{2df-clust-lf} from the 2dF galaxy redshift survey.
In the light of our model, and of the lack of strong dips in the LFs
of high $L_X$ groups, what are we to make of the presence of such dips
in a minority of rich clusters, such as the Coma cluster
(Fig.~\ref{fig:comalf})? Clearly if one were to create a cluster by
merging a set of X-ray dim groups, such as those in our sample, the
result would be a cluster with a dip in the LF as seen in Coma. It
therefore seems likely that the diversity in the LFs of richer
clusters reflects a diversity in their formation history. Close study
\citep{white93} has shown Coma to be a rather unusual cluster,
which shows evidence for a turbulent recent history involving multiple
subcluster mergers. We speculate that clusters with strong LF dips may
have been assembled primarily from multiple mergers of low velocity
dispersion groups.

The results presented in this paper
suggest that the low-$L_X$ groups are sites of
recent dynamical evolution. Their high-$L_X$ counterparts, may
have gone through this stage in the past with the luminosity function
evolving such that the intermediate-luminosity galaxies are gradually
depleted and one or two bright central galaxies form from their
merger. The intermediate-luminosity bins may then be re-filled through
filamentary infall.

This picture of galaxy evolution evolution leads to a definite
prediction.  If groups of low velocity dispersion are indeed systems
that are undergoing rapid dynamical evolution, the stellar populations
in their galaxies would be significantly younger than those in
high-$\sigma$ groups in case of dissipative merging.  This can be
observationally verified using sophisticated techniques to determine
age without significant ambiguities from the effects of metal
abundance \citep[e.g.][]{nolan03,tf02}.

\begin{figure}
\begin{center}
\epsfig{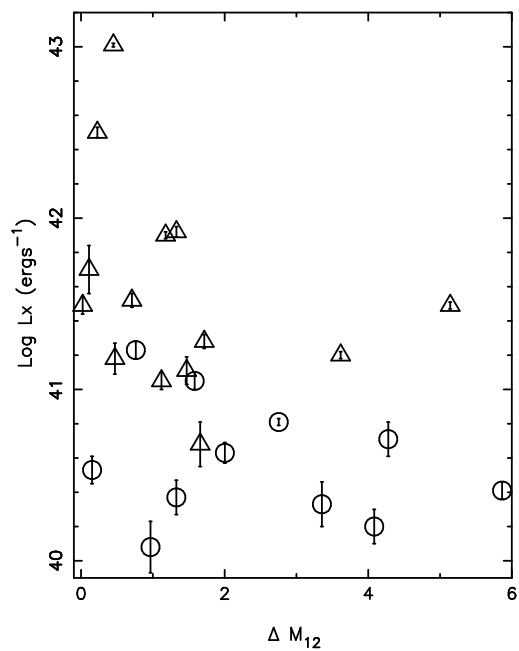}
\caption{The X-ray luminosity of our sample groups, as a function of 
the difference $\Delta M_B$ 
between the absolute blue magnitudes of the brightest
and second brightest galaxies.
Groups with detected X-ray emission associated with the group, and
not just with individual galaxies (see Table~1) are plotted as circles.
Groups with lower $L_X$ have systematically higher $\Delta M_B$.
\label{fig:Lxmm2}}
\end{center}
\end{figure}

The above scenario seems to be supported by Fig.~\ref{fig:Lxmm2},
which shows the relation between the X-ray luminosity of our groups
and the difference in magnitude between their brightest and second
brightest galaxies. The X-ray bright groups have several galaxies of
comparable luminosity (and mass) at the bright end, being the
end-products of earlier mergers on smaller scales in sub-groups that
were incorporated in the virialised systems we observe today.

\section*{Acknowledgements}
Thanks to Ale Terlevich for his involvement in observing and data
reduction, John Osmond for his work on the GEMS catalogue and Habib
Khosroshahi for useful discussions. VKP and PG would like to thank the
director of STScI for financial support through the Director's
Discretionary Research Fund.  This
paper is based upon observations collected at the Issac Newton
Telescope, La Palma (Observing Programme I/00A/23) and observations
gathered at the European Southern Observatory, Chile (Observing
Programme 67.A-0252(A)).  This research has made use of the NASA/IPAC
Extragalactic Database (NED) which is operated by the Jet Propulsion
Laboratory, California Institute of Technology, under contract with
the National Aeronautics and Space Administration.

\bsp

\label{lastpage}
\end{document}